\documentclass[cameraready]{Interspeech}

\usepackage{xcolor}

\usepackage{comment}
\usepackage{booktabs}
\usepackage{todonotes} 
\usepackage{textcomp}
\usepackage{amsmath}
\usepackage{bbm}
\usepackage{arabtex}
\usepackage{utf8}
\setcode{utf8}
\usepackage{subcaption}

\usepackage{xcolor}
\usepackage{soul}
\usepackage{makecell}

\usepackage{graphicx}
\definecolor{lightblue}{rgb}{.50,.95,1}
\definecolor{tri}{rgb}{.25,.88,.82}
\definecolor{lilac}{rgb}{0.85,0.64,0.85}

\newcommand{\dataset}{\emph{WASIL}}

\newcommand{\data}{\emph{WASIL}}

\definecolor{tri}{rgb}{.25,.88,.82}
\definecolor{lilac}{rgb}{0.85,0.64,0.85}
\definecolor{lightblue}{rgb}{.50,.95,1}

\usepackage{listings}
\usepackage{xcolor}

\lstdefinestyle{promptstyle}{
  basicstyle=\ttfamily\footnotesize,
  breaklines=true,
  breakatwhitespace=true,
  columns=fullflexible,
  keepspaces=true,
  showstringspaces=false,
  framerule=0.4pt,
  xleftmargin=0.6em,
  xrightmargin=0.6em,
  aboveskip=0.6em,
  belowskip=0.6em
}

\title{WASIL: In-the-Wild Arabic Spoken Interactions with LLMs}

\author[affiliation={},orcid=0000-0001-6698-2842]{Zien Sheikh}{Ali}
\author[affiliation={},orcid=0000-0003-4828-6098]{Hamdy}{Mubarak}
\author[affiliation={}]{Soon-Gyo}{Jung}
\author[affiliation={},orcid=0009-0006-0388-4712]{Hunzalah Hassan}{Bhatti}
\author[affiliation={},orcid=0000-0001-7172-1997]{Firoj}{Alam}
\author[affiliation={},orcid=0000-0002-1331-2543]{Shammur Absar}{Chowdhury}

\address{
    Qatar Computing Research Institute, Qatar
}

\email{shchowdhury@hbku.edu.qa}


\begin{document}

\maketitle

\begin{abstract}
Large Language Models (LLMs) voice assistants are commonly built as cascaded Automatic Speech recognition (ASR) → LLM systems, where recognition errors can distort user intent. Dislikes may also arise from ambiguous, out-of-domain, or non-request turns, making it hard to isolate ASR effects. We release \dataset{},\footnote{WASIL \<واصل> denotes connection or linking in Arabic.} in-the-wild Arabic spoken interaction prompts with audio, ASR hypotheses, assistant responses, and explicit like/dislike feedback, covering Modern Standard Arabic (MSA) and four major dialects with their labels. We provide low-cost gold transcripts via multi-ASR agreement–guided post-editing and annotate answerability (answerable, ambiguous/needs-clarification, unsupported, not-a-request/noise) to separate intrinsic unanswerability from ASR-induced degradation. Finally, we describe scalable reference-free evaluation of responses from ASR vs. gold transcripts using multi-judge LLM scoring. 
\end{abstract}

\section{Introduction}

Large language models (LLMs) are increasingly embedded in everyday applications, supporting both text and speech interaction and enabling open-domain conversational assistants beyond intent--slot pipelines \cite{wang2025voiceassistant_eval,hou2025sova_bench}. In many practical systems, speech interaction is implemented as a cascade in which automatic speech recognition (ASR) first converts audio to text, and the resulting transcript is then processed by an LLM \cite{billa2026cascade_equivalence}. This design introduces a key evaluation challenge because user dissatisfaction may stem from multiple confounded sources, including \textit{(i)} ASR errors that distort user intent \cite{kubis-etal-2023-back}, \textit{(ii)} intrinsically unclear or effectively unanswerable user turns that require clarification or repair \cite{galbraith2024dialogue_repair,men2025reject_or_not}, and \textit{(iii)} limitations of the downstream LLM itself \cite{chen2024voicebench}. Separating these factors is difficult without the original audio, alternative transcriptions, and fine-grained labels that isolate error sources across the pipeline.

Recent ASR research has therefore pushed evaluation beyond word error rate (WER) toward measures that better reflect downstream conversational success. Semantic-oriented metrics such as semantic distance (SemDist) quantify meaning preservation and, in many settings, align more closely with spoken language understanding performance than surface-form matching \cite{kim21e_interspeech}. Such metrics are particularly relevant when transcripts feed directly into an LLM, where small wording changes can yield large differences in interpretation.

Complementary work on significant error detection further suggests that only a subset of recognition errors are truly damaging for assistants \cite{harvill2024sasred}. For LLM-based voice assistants, Liu et al.\ also show that transcript quality can materially affect generated responses and propose downstream-oriented ASR evaluation approaches that account for this sensitivity \cite{liu24c_interspeech}. Despite this progress, much of the existing evidence is based on curated prompts, controlled settings, or predominantly English data. This leaves open which factors most strongly drive user dislike in real spoken interactions, especially in Arabic where dialect diversity and non-standard orthography can increase transcription ambiguity \cite{talafha-etal-2024-casablanca,ali2021connecting}.

In parallel, large-scale interaction logs and preference signals have accelerated evaluation and modeling for text chat assistants \cite{zhao2024wildchat}. Comparable resources for spoken interaction with LLM-based assistants remain limited~\cite{alam2025everydaymmqa,ali2026menaspeechbankreferencevoicebank}. Existing benchmarks for LLM-based voice assistants, including VoiceBench, provide valuable task-driven tests but do not typically reflect naturally occurring user sessions with explicit feedback \cite{chen2024voicebench}. Public Arabic datasets have largely prioritized ASR development and dialect coverage \cite{talafha-etal-2024-casablanca}, rather than end-to-end spoken interaction with LLMs paired with user reactions and labels that separate ASR errors from user-turn ambiguity.

We address this gap with \data{}, a dataset of in-the-wild Arabic spoken interactions with an LLM-based assistant. We collect $\sim$9K turns (9,304 turns from 93 users),\footnote{Note that, our evaluation and analysis in this study was based on 2,573 turns.} spanning multiple dialects and countries, and including explicit user feedback on assistant responses, including like or dislike signals and scalar scores. 


\noindent
Our contributions are:
\begin{itemize}
  \item We release, to our knowledge, the first dataset of in-the-wild \emph{Arabic spoken} LLM interaction prompts with explicit user feedback, post-edited gold transcripts, and intrinsic answerability labels. We release the \dataset{} test dataset publicly to the community.\footnote{\url{https://huggingface.co/datasets/QCRI/WASIL}} 
  \item We propose a low-cost reference creation strategy that uses multi-ASR agreement as a proxy for transcription reliability, reducing human effort relative to transcribing from scratch.
  \item We introduce an annotation schema for intrinsic answerability and clarity, and analyze its association with user dislikes, which helps separate failures driven by transcription issues from those driven by response content.
  \item Using the gold transcripts, we benchmark multiple LLMs and conduct reference-free response evaluation with multi-judge LLM-as-a-judge protocols, enabling comparison without gold answers.
\end{itemize}

\section{Related Work}
\label{sec:related}

\subsection{Interaction Datasets}

Large-scale logs of human--assistant interactions have enabled empirical analysis of failure modes and preference learning for text-based assistants. WildChat collects one million real ChatGPT interaction logs \cite{zhao2024wildchat}. Chatbot Arena provides pairwise human preferences and an Elo-style ranking framework for LLM evaluation \cite{chiang2024chatbotarena}, and MT-Bench analyzes the agreement between LLM judges and human preferences \cite{zheng2023judging}. These resources motivate our focus on naturally occurring data and direct user feedback. In contrast, \textbf{our setting is spoken}, cascaded (ASR$\rightarrow$LLM), and Arabic, introducing recognition errors and dialectal variation.

\subsection{Evaluating ASR for LLM-based Systems}
WER is widely used as the standard metric, however, it can be poorly aligned with downstream task quality measurement. SemDist measures semantic discrepancy in an embedding space and better captures semantic equivalence for spoken language understanding tasks \cite{kim21e_interspeech}. Harvill et al.\ propose detecting \emph{significant} ASR errors that are important for conversational voice assistants, reflecting the observation that not all errors equally harm downstream behavior \cite{harvill2024sasred}. For LLM-based voice assistants specifically, Liu et al.\ study how ASR performance impacts an LLM’s generated responses and propose evaluation methods targeting the downstream effect \cite{liu24c_interspeech}. Our work complements these lines by connecting transcript quality to \emph{real user} like/dislike feedback 
so that ``downstream failures'' are not conflated with unclear user turns.

\subsection{Answerability in Spoken Conversation}
Spoken dialogue systems have long studied non-understanding events, rejection, and recovery strategies, showing these errors strongly affect perceived quality \cite{bohus2005nonunderstanding}. Modern assistants additionally face out-of-domain utterances that fall outside supported capabilities; Tur et al.\ frames this as detecting rejected but assistant-directed out-of-domain requests \cite{tur2014detecting}. A complementary perspective is deciding when to ask clarifying questions. Rahmani et al.\ survey datasets and evaluation practices for clarification in conversational systems \cite{rahmani-etal-2023-survey}. These works motivate our ``intrinsic answerability'' labels as a lightweight way to separate turns that are answerable from those that require clarification, are out-of-scope.

\subsection{Low-cost Reference Transcription}
Creating high-quality reference transcripts for in-the-wild speech is expensive, especially for dialectal Arabic where orthographic variation yields multiple acceptable transcriptions. System combination methods such as ROVER exploit agreement across ASR hypotheses to reduce errors \cite{fiscus1997rover}. In dialectal Arabic ASR, Ali et al.\ advocate multi-reference evaluation to account for transcription variability \cite{ali-etal-2015-multi}, and Wray et al.\ discuss best practices for crowdsourcing dialectal Arabic speech transcription \cite{wray-etal-2015-best}. 
More recently, multi-ASR fusion combined with LLM-based correction has been explored for producing pseudo-labels for ASR training \cite{prakash25_interspeech}, supporting our premise that cross-system agreement can serve as a proxy for transcript reliability. Our dataset and analysis instantiate these ideas in a real spoken LLM interaction setting and quantify their implications for downstream response quality. 


\section{Datasets}

\subsection{Data Collection}\label{sec:data_coll}
\begin{figure}[!tbh]
    \centering
    \includegraphics[width=\columnwidth]{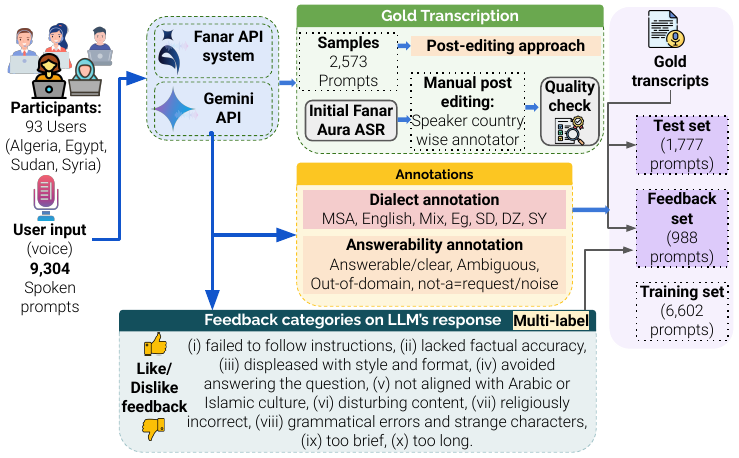}
    \caption{\textbf{WASIL} dataset development process, from multi-national spoken prompts collection and cascaded model inference to multi-layer human annotation.}
    \label{fig:dataset_pipeline}
    \vspace{-0.2cm}
\end{figure}

In Figure \ref{fig:dataset_pipeline}, we present \dataset{} dataset development process. For data collection, we recruited 93 users to interact with an Arabic-centric ASR$\rightarrow$LLM system. For both tasks, we used the publicly available Fanar APIs\footnote{\url{https://api.fanar.qa/docs}} \cite{fanar2024}. The same user recordings were also processed with an alternative pipeline that uses Gemini~\cite{team2023gemini} for both ASR and LLM inference. As part of cascaded system we have also used ALLaM~\cite{bari2024allam} LLM. 

Participants were recruited from four Arab countries, Algeria, Egypt, Sudan, and Syria. Data were collected over nine days using daily topic prompts that encouraged diverse interactions, including open discussion (any topic), follow-up questions, creative writing, scientific questions, safety-related queries, and Arabic culture. Users were not restricted to interact only in Modern Standard Arabic (MSA), so the recordings include a range of Arabic dialect varieties. In total, we collected \textbf{9,304 spoken prompts}, also referred to as spoken utterances.


Users were recruited through a third-party company and compensated for their time. During platform sign-up, participants provided informed consent for their interaction data to be used for further use.

In addition to the spoken interaction, they have also labeled responses as \textbf{like} and \textbf{dislike}. For the dislike cases, a set of responses has also been labeled feedback categories, which includes: (i) \textit{failed to follow instructions}, (ii) \textit{lacked factual accuracy}, (iii) \textit{displeased with style and format}, (iv) \textit{avoided answering the question}, (v) \textit{not aligned with Arabic or Islamic culture}, (vi) \textit{disturbing content}, (vii) \textit{religiously incorrect}, (viii) \textit{grammatical errors and strange characters}, (ix) \textit{too brief}, (x) \textit{too long}. 
  
To interpret dislikes, we use the above multi-label failure taxonomy that keeps common error sources separate instead of collapsing them into a single rating, which follows multi-metric evaluation practice in HELM \cite{liang2022helm}. The labels cover instruction adherence and answer completeness, which are central targets of instruction tuning and RLHF \cite{ouyang2022instructgpt} and are assessed by verifiable instruction-following tests such as IFEval \cite{zhou2023ifeval}. We annotate factual inaccuracy because hallucinations are a persistent failure mode captured by truthfulness-style benchmarks \cite{lin2021truthfulqa}. We also distinguish inappropriate content from avoidance and over-refusal to capture the helpfulness-harmlessness trade-off emphasized in safety alignment work \cite{bai2022constitutional} and studied in over-refusal evaluation such as OR-Bench \cite{cui2024orbench}. Usability issues are split into style and format, fluency artifacts such as strange characters, and length problems because presentation and readability are known to strongly influence human judgments \cite{fabbri2021summeval}. Finally, we include Arabic 
cultural and religious labels to reflect documented cultural value bias in LLM outputs \cite{tao2024cultural} and the need for explicit evaluation of religious-value alignment.

\subsection{Gold transcript}
We selected 2,573 turns for gold transcription. Of these, 1,777 turns were randomly sampled to form the \textbf{\textit{test set}} used throughout the paper. We also collected a \textbf{\textit{feedback set}}, consisting of 988 turns from the full dataset in which users provided explicit feedback on assistant responses. We use this set for feedback-category analysis.\footnote{Note that the test and feedback sets are not mutually exclusive and partially overlap.}  In addition to the gold-transcribed data, we plan to release the remaining turns for analysis and model training in future work.

To construct the gold transcripts, we adopted a post-editing approach. As mentioned earlier, we used Fanar’s Aura ASR (Fanar-Aura-STT-1)\footnote{\url{https://api.fanar.qa/docs}} to generate the initial transcripts. We then manually post-edited all 2,573 transcripts. 

To ensure dialectal and linguistic accuracy, each utterance was assigned to an annotator based on the speaker's country of origin. For example, utterances produced by Egyptian speakers were reviewed by Egyptian annotators. Annotators were provided with standardized annotation guidelines to ensure consistency. To ensure the quality of the transcription an expert from our team randomly checked the transcription and revised the transcriptions in a second iteration.

\subsection{Dialect Annotation}
As noted in Section~\ref{sec:data_coll}, users were free to speak in any language or dialect. We therefore added a second annotation step to label the dialect of each utterance. The same group of annotators listened to the audio and labeled each utterance as MSA, English, Mix (dialect--English code-switching), or dialectal Arabic. Each utterance was labeled by one annotator and later validated by our team for consistency. We adopt this design because dialect identification from speech is usually consistent for trained native listeners, and prior work reports very high agreement for similar dialect identification labeling \cite{jones2024maglic}. We then add a validation step to handle borderline cases and to ensure consistent decisions across annotators, which aligns with common multi-stage corpus annotation practice \cite{hamed2024zaebucspoken,artstein2008intercoder}.


\subsection{Answerability Annotation} 
We annotate intrinsic answerability for each user utterance on the reference transcripts, with the aim of separating prompts that a text-only assistant can address immediately from those that require a different interaction strategy. We use four mutually exclusive labels, \textit{Answerable/Clear}, \textit{Ambiguous/Needs-Clarification}, \textit{Out-of-Domain/Unsupported}, and \textit{Not-a-Request/Backchannel/Noise}. This choice reflects common system behaviors reported in prior dialogue research. Underspecified turns and unclear referents are naturally handled through clarification, which is a standard grounding and repair mechanism in spoken dialogue systems \cite{akasaki2024ambiguous}. Requests that are well formed but outside the supported intent set or required capabilities are separated as \textit{Out-of-Domain/Unsupported}, consistent with work on out-of-scope and out-of-domain detection in task-oriented assistants \cite{zhan2021oos, lang2024ood}. Finally, backchannels, social tokens, and fragments are grouped as \textit{Not-a-Request/Backchannel/Noise} since they do not introduce new task content, aligning with dialogue act standards and corpus guidelines that model these phenomena separately \cite{bunt2020iso}. Each transcript is independently labeled by three annotators familiar with the target dialect, following the annotation guideline with examples to support consistent annotation decisions.

We used three annotators to mitigate subjectivity and reduce individual annotator bias in this inherently subjective task.  The final label was determined by majority vote. To quantify inter-annotator reliability, we compute agreement using \textbf{Gwet's AC1}~\cite{gwet2008computing} for nominal categories. AC1 provides a stable chance-corrected estimate under high agreement and skewed label distributions, and it is less sensitive to prevalence effects than commonly used alternatives such as Cohen's $\kappa$ or Krippendorff's $\alpha$. The annotation agreement is AC1=0.65, indicating moderate-to-substantial agreement. 

\subsection{Annotation Task Setup}
For each annotation task, we provided clear written guidelines and initial instructions with examples. To facilitate the annotators we provided Arabic version of the annotation guidelines. The full guidelines are included in the supplemental material. All annotations were completed on our in-house annotation platform. Annotators were recruited through a third-party company and compensated at a standard hourly rate, following the same procedure as in spoken-prompt collection. All annotators were professionals fluent in both Arabic and English and held at least a bachelor's degree. Each annotator signed a non-disclosure agreement to meet institutional requirements and to permit downstream use of the data. For each annotation task, we ensured annotation quality through periodic checks of randomly sampled items and provided feedback to the annotators.  
 

\subsection{Analysis}
\subsubsection{What Percentage of the Transcripts Were Post-Edited?}
In Table~\ref{tab:postedit_vs_asr}, we report the performances of gold transcripts vs. different ASR systems. The average WER of 0.190 and the semantic similarity of 0.917 for \textbf{Fanar} indicate that ASR hypotheses often benefit from correction, supporting post-editing as a reliable approach for developing gold transcripts. Note that for semantic similarity we extract embeddings and compute cosine similarity. For embedding extraction, we used the \textit{paraphrase-multilingual-mpnet-base-v2} model from the Sentence-Transformers library.\footnote{\url{https://huggingface.co/sentence-transformers/}}

\noindent
\subsubsection{ASR Performance Across Systems} 
As presented in Table~\ref{tab:postedit_vs_asr}, overall, Gemini and Fanar yield identical average WER (0.18), indicating a comparable level of word-errors. In contrast, Gemini achieves higher semantic agreement (0.92 vs.\ 0.89 cosine similarity), suggesting that its hypotheses more often preserve the intended meaning even when surface forms differ. 
\begin{table}[t]
\centering
\setlength{\tabcolsep}{1pt} 
\scalebox{0.7}{
\begin{tabular}{llcccc}
\toprule
\textbf{Dialect} & 
\makecell{\textbf{\#Prompts} \\ \textbf{(Utter.)}} & 
\multicolumn{2}{c}{\textbf{Avg. WER}} & 
\multicolumn{2}{c}{\textbf{Avg. Cosine Sim.}} \\
\cmidrule(lr){3-4}\cmidrule(lr){5-6}
{} & {} & {\textbf{Fanar}} & {\textbf{Gemini}} & {\textbf{Fanar}} & {\textbf{Gemini}} \\
\midrule
MSA & 979 & 0.21 & 0.16 & 0.92 & 0.97 \\
EG & 868 & 0.12 & 0.10 & 0.95 & 0.98 \\
MIX & 239 & 0.23 & 0.25 & 0.75 & 0.80 \\
SD & 234 & 0.22 & 0.14 & 0.91 & 0.95 \\
DZ & 143 & 0.12 & 0.19 & 0.94 & 0.95 \\
EN & 55 & 0.28 & 0.27 & 0.82 & 0.85 \\
SY & 55 & 0.09 & 0.15 & 0.95 & 0.97 \\ 
\midrule
\textbf{Avg.} & \textbf{2,573} & \textbf{0.18}& \textbf{0.18} & \textbf{0.89} & \textbf{0.92} \\
\bottomrule
\end{tabular}
}
\caption{Agreement scores (WER and cosine similarity) between post-edited transcripts and outputs from two ASR systems, grouped and sorted by spoken-prompts. The Avg. row reports a prompt-weighted average across dialects.}
\label{tab:postedit_vs_asr}
\vspace{-0.4cm}
\end{table}


\noindent
\subsubsection{Dialect-wise Differences.} 

Agreement varies substantially by dialect. Egyptian Arabic and Syrian Arabic show the strongest performance for both systems, with low WER (0.09-0.15) and high semantic similarity (0.95-0.98). MSA is moderately harder in our setting, with higher WER (0.16-0.21) and slightly lower similarity than EG and SY. One contributing factor is Arabic diglossia. Everyday speech is typically produced in regional dialects, while MSA is a formal variety learned primarily through schooling. In practice, spoken requests that are intended to be MSA often include dialectal pronunciations, lexical choices, or light mixing, which increases acoustic and lexical mismatch for ASR \cite{elmahdy10_interspeech}. Sudanese Arabic shows the largest degradation among the major Arabic dialects, with WER in the 0.14-0.22 range and cosine similarity around 0.91-0.95, reflecting more frequent recognition distortions. Mixed-dialect utterances are also challenging, where WER rises to 0.25 (Gemini) and similarity drops to 0.75 (Fanar). This pattern is consistent with dialectal code-switching and non-standard lexical variation \cite{chowdhury21_interspeech}. Finally, English utterances exhibit relatively low semantic similarity (0.82-0.85) despite higher WER across systems (0.27-0.28), suggesting that errors in English are more likely to change meaning rather than preserve intent.

\noindent
\subsubsection{Is Fully Manual Transcription Necessary, or Is Post-Editing Sufficient?}
We transcribed 100 utterances from scratch and compared them with the corresponding post-edited transcripts. Using the fully manual transcripts as reference, the post-edited versions show strong agreement (WER = 0.07) and nearly identical meaning (cosine similarity = 0.98). This suggests that post-editing is sufficient to produce reliable reference transcriptions. This observation is consistent with recent work that applies LLM-based postprocessing to correct ASR outputs and reports corrected transcripts that are close to ground truth and usable as high-quality labels for training \cite{prakash25_interspeech}.

\noindent
\subsubsection{Cost Optimal Approach for Gold Transcription.} 
\begin{figure}[h]
    \centering
    \includegraphics[width=0.9\columnwidth]{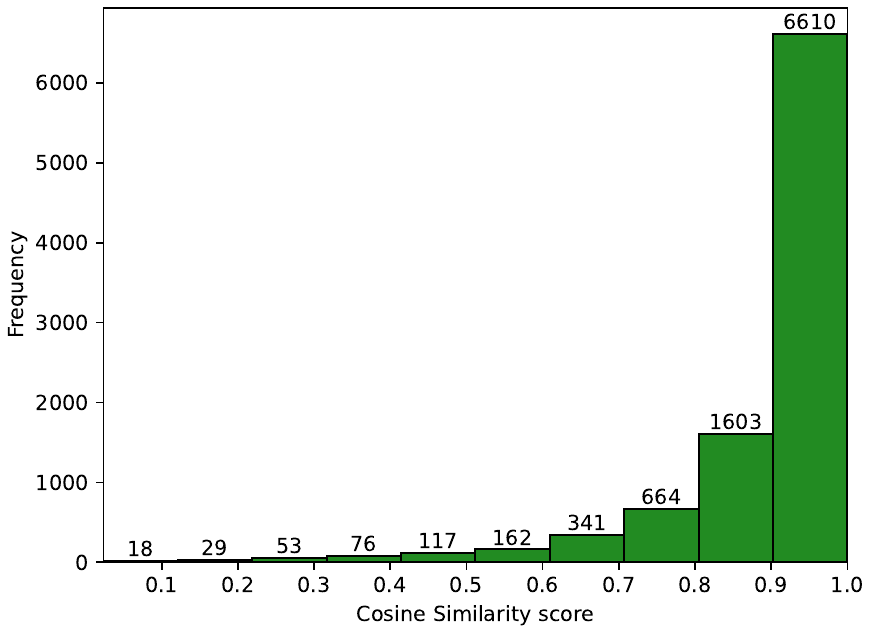}
    \caption{The distribution of cosine similarity scores between Fanar and Gemini transcriptions on the whole dataset (9,304).}
    \label{fig:asr_sim_histogram}
    \vspace{-0.3cm}
\end{figure}

For each utterance we collect multiple ASR hypotheses from distinct ASR systems such as Fanar and Gemini. We compute an agreement score i.e., pairwise normalized edit distance and use it to drive a cost-aware approach. In line with prior work such as \cite{prakash25_interspeech}, our hypothesis is that high-agreement utterances are accepted with minimal edits, while low-agreement utterances are prioritized for post-editing. 
Figure \ref{fig:asr_sim_histogram} presents the distribution of cosine similarity scores between Fanar and Gemini transcriptions. A high level of agreement is observed, with 68\% of utterances exceeding 0.90 similarity, supporting the reliability of the proposed hypothesis. Assuming only utterances below 0.90 require post-editing, manual effort could be reduced by 68\%. 

\begin{figure}[h]
    \centering
    \includegraphics[width=\columnwidth]{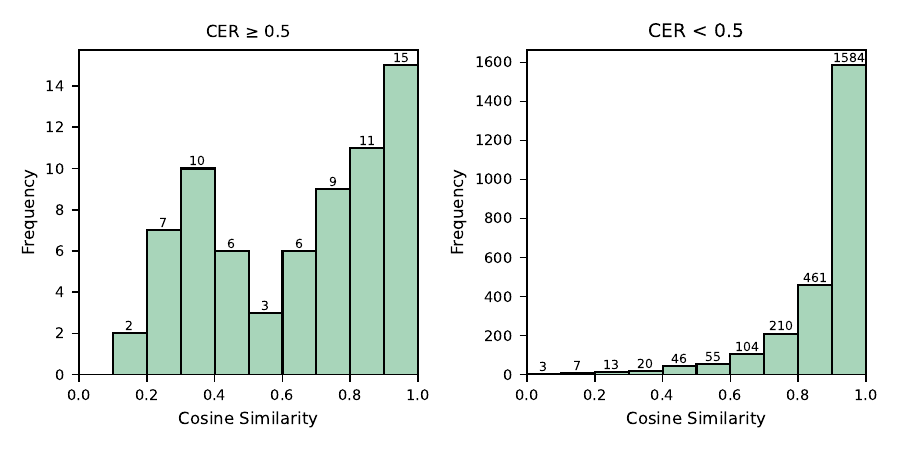}
    \caption{Cosine similarity between Fanar and Gemini transcriptions for post-edited utterances, split into higher CER cases that required more edits (left) and lower CER cases that required fewer edits (right).}
    \label{fig:asr_sim_CER}
    \vspace{-0.3cm}
\end{figure}

\begin{figure}[h]
    \centering
    \includegraphics[width=\columnwidth]{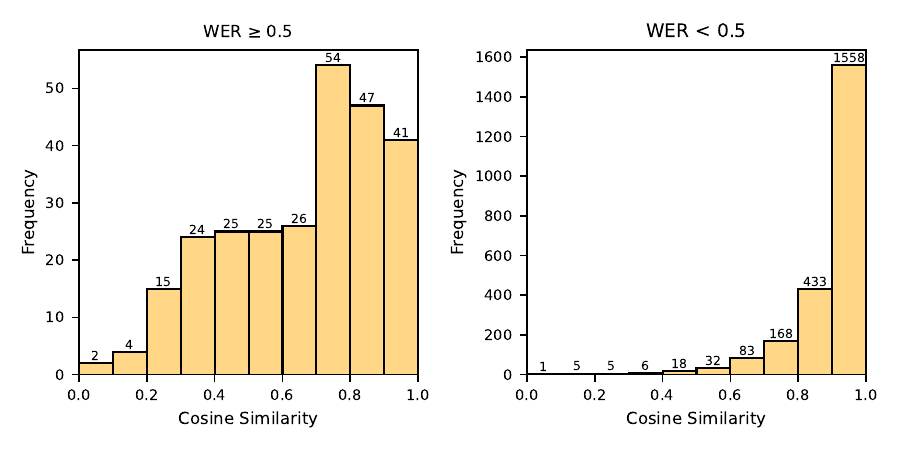}
    \caption{Cosine similarity between Fanar and Gemini transcriptions for post-edited utterances, split into higher WER cases that required more edits (left) and lower WER cases that required fewer edits (right).}
    \label{fig:asr_sim_WER}
    \vspace{-0.3cm}
\end{figure}

To further validate this hypothesis, we conduct an additional analysis using the gold transcriptions. We compute word-level (WER) and character-level (CER) edit distances between the gold transcriptions and the Fanar ASR outputs to quantify the actual post-editing effort required. Figures \ref{fig:asr_sim_CER} and \ref{fig:asr_sim_WER} present these results. We partition the data into two groups: prompts requiring minimal edits (low WER/CER) and prompts requiring substantial edits (high WER/CER). Consistent with our hypothesis, prompts requiring minimal edits exhibit higher inter-ASR agreement, whereas prompts requiring substantial edits show lower agreement, confirming that ASR agreement is a useful indicator of post-editing effort.

\noindent
\subsubsection{ASR Errors vs. Response Dislikeness}

\begin{figure}[h]
    \centering
    \includegraphics[width=\columnwidth]{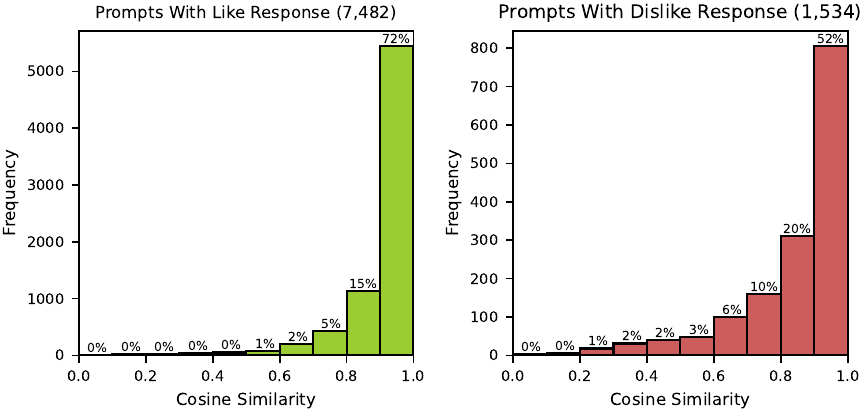}
    \caption{Distribution of cosine similarity scores between Fanar and Gemini transcriptions for spoken prompts categorized by like (left) and dislike (right) reactions.}
    \label{fig:asr_sim_likes_dislikes}
    \vspace{-0.3cm}
\end{figure}

We investigate the potential reasons behind dislike reactions by examining the relationship between ASR similarity scores and user feedback. Figure \ref{fig:asr_sim_likes_dislikes} presents the distribution of cosine similarity scores between Fanar and Gemini transcriptions for prompts where the response received a like reaction (left) and a dislike reaction (right).
Among prompts with a like reaction, 72\% exhibit high similarity ($\geq 0.9$), suggesting good transcription quality. In contrast, only 52\% of prompts with a dislike reaction achieve a similarity score $\geq 0.9$, while a larger proportion showing lower similarity scores.

\begin{figure}[h]
     \centering
     \begin{subfigure}[b]{\columnwidth}
         \centering
         \includegraphics[width=\columnwidth]{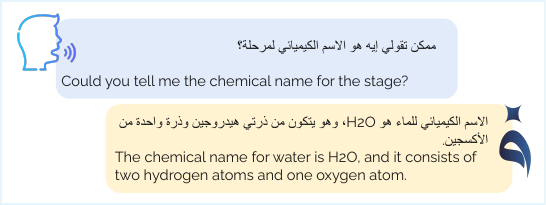}
        \caption{Example of an ASR transcription error leading to an incorrect response. The user prompt is answerable, but the incorrect transcription results in an incorrect system response.}
        \label{fig:asr_errors_wrong}
     \end{subfigure}
     \begin{subfigure}[b]{\columnwidth}
        \includegraphics[width=\columnwidth]{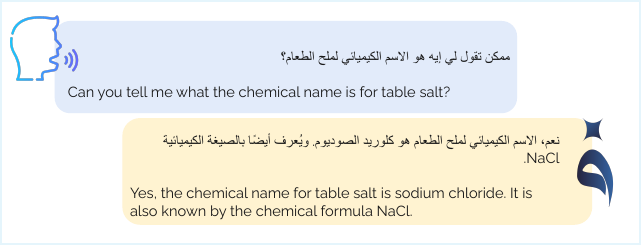}
        \caption{Correct response after providing the gold transcription. When the correct transcription is used instead of the ASR output, the system produces the correct answer.}
        \label{fig:asr_errors_corrected}
     \end{subfigure}     
     \begin{subfigure}[b]{\columnwidth}
    \centering
    \includegraphics[width=\columnwidth]{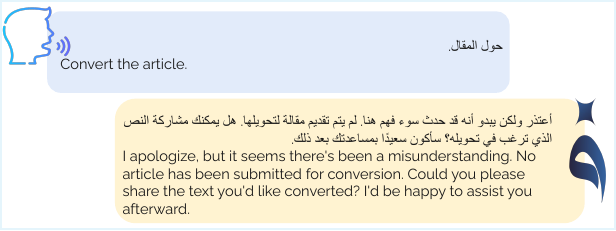}
    \caption{Example of a prompt that received a dislike reaction despite correct transcription. The prompt lacks sufficient detail.}
    \label{fig:prompt_needs_clarification}
    \vspace{-0.3cm}
     \end{subfigure}
     \caption{Examples of different prompts causing user dislike reactions}
     \label{fig:three_stacked_images}
     \vspace{0.4cm}
\end{figure}

We present examples of prompts that received dislike feedback from users. Figure \ref{fig:asr_errors_wrong} shows a case where the user prompt is answerable, but a transcription error leads to an incorrect transcription and consequently an incorrect response. In Figure \ref{fig:asr_errors_corrected}, we provide the gold transcription to Fanar, which results in the correct answer.

We also present an example of a prompt without transcription errors that still received a dislike reaction (Figure \ref{fig:prompt_needs_clarification}). In this case, the prompt is incomplete and lacks sufficient detail, which leads to the system to request additional information instead of providing an answer.

\noindent
\subsubsection{Feedback Category Analysis}
\label{sec:meta_error_taxonomy}

\noindent
Human feedback on LLM outputs is often multi-faceted. A single response can be unhelpful (e.g., fails to follow instructions), incorrect (factual or domain errors), unsafe (disturbing/toxic), poorly written (grammar/formatting), or culturally misaligned. To make our analysis interpretable and comparable to prior work, we map our fine-grained, human-labeled error categories into a small set of \emph{meta-dimensions} that are widely used in alignment and NLG evaluation.


Let $y_i$ denote the set of atomic error labels assigned to instance $i$ (multi-label).
We define a deterministic mapping $m(\cdot)$ from each atomic label to one or more meta-feedback, and compute a meta-profile
$z_i = \bigcup_{\ell \in y_i} m(\ell)$.
Table~\ref{tab:meta_taxonomy} summarizes the mapping used in our experiments.
This decision matches annotator intent in settings where ``religiously wrong'' denotes incorrect religious facts or rulings.

\begin{table}[t]
\centering
\setlength{\tabcolsep}{1pt} 
\scalebox{0.7}{
\begin{tabular}{@{}p{0.3\linewidth}p{0.55\linewidth}p{0.55\linewidth}@{}}
\toprule
\textbf{Meta-dimension} & \textbf{Rationale} & \textbf{Atomic labels} \\
\midrule
\textbf{A. Helpfulness / task success} &
Instruction following and successful task completion \cite{ouyang-etal-2022-training}. &
\texttt{failed to follow instructions}, \texttt{avoided answering questions}, \texttt{too brief}, \texttt{too long}. \\
\addlinespace
\textbf{B. Correctness / truthfulness} &
Factual and domain correctness; ``honesty'' is often operationalized via truthfulness benchmarks \cite{lin2021truthfulqa}. &
\texttt{lacked factual accuracy}, \texttt{religiously wrong}. \\
\addlinespace
\textbf{C. Safety / harmlessness} &
Avoiding harmful/inappropriate content and toxic degeneration \cite{gehman-etal-2020-realtoxicityprompts}. &
\texttt{content was disturbing}. \\
\addlinespace
\textbf{D. Communication quality} &
Surface-form quality (grammar, readability, formatting) is commonly separated from task success in NLG evaluation \cite{gatt-krahmer-2018-survey}. &
\texttt{displeased with style and format}, \texttt{grammatically incorrect}, \texttt{full of grammatical errors and strange letters}. \\
\addlinespace
\textbf{E. Cultural \& religious alignment} &
Capturing mismatches with local norms/values; Arabic/Islamic cultural competence is increasingly evaluated explicitly \cite{naous-etal-2024-beer,alwajih-etal-2025-palmx}. &
\texttt{not aligned with arabic or islamic culture}. \\
\bottomrule
\end{tabular}
}
\caption{Mapping of meta-feedback used to summarize fine-grained labels. As labels are multi-label, a single instance can be assigned to multiple meta-feedback dimensions.}
\label{tab:meta_taxonomy}
\vspace{-0.3cm}
\end{table}

\begin{table}[t]
\centering
\setlength{\tabcolsep}{3pt}
\scalebox{0.7}{
\begin{tabular}{p{9.0cm}rr}
\toprule
\textbf{Meta feedback} & \textbf{Count} & \textbf{\%} \\
\midrule
A Helpfulness/Task success & 373 & 37.8 \\
B Correctness/Truthfulness & 239 & 24.2 \\
D Communication quality & 114 & 11.5 \\
A Helpfulness/Task success; D Communication quality & 30 & 3.0 \\
B Correctness/Truthfulness; D Communication quality & 29 & 2.9 \\
A Helpfulness/Task success; B Correctness/Truthfulness; D Communication quality & 25 & 2.5 \\
E Cultural alignment & 18 & 1.8 \\
C Safety/Harmlessness & 13 & 1.3 \\
B Correctness/Truthfulness; A Helpfulness/Task success; D Communication quality & 12 & 1.2 \\
A Helpfulness/Task success; A Helpfulness/Task success & 11 & 1.1 \\
C Safety/Harmlessness; E Cultural alignment & 10 & 1.0 \\
D Communication quality; E Cultural alignment & 10 & 1.0 \\
B Correctness/Truthfulness; B Correctness/Truthfulness; D Communication quality & 10 & 1.0 \\
E Cultural alignment; D Communication quality & 9 & 0.9 \\
D Communication quality; A Helpfulness/Task success & 6 & 0.6 \\
A Helpfulness/Task success; D Communication quality & 6 & 0.6 \\
\bottomrule
\end{tabular}
}
\caption{Counts of mutually exclusive mapped feedback meta-profiles after mapping atomic labels into meta-clusters (A--E). $N=988$ feedback instances.}
\label{tab:meta-profile-counts}
\vspace{-0.3cm}
\end{table}

\begin{table}[!tbh]
\centering
\setlength{\tabcolsep}{3pt}
\scalebox{0.85}{
\begin{tabular}{llrr}
\toprule
\textbf{Cluster} & \textbf{Meaning} & \textbf{Count} & \% \\
\midrule
A & Helpfulness/Task success  & 522 & 40.7 \\
B & Correctness/Truthfulness  & 384 & 29.9 \\
D & Communication quality     & 263 & 20.5 \\
E & Cultural alignment        & 68  & 5.3  \\
C & Safety/Harmlessness       & 46  & 3.6  \\
\bottomrule
\end{tabular}
}
\caption{Marginal prevalence of each meta-cluster (non-mutually exclusive; $N=988$).}
\label{tab:meta-cluster-marginals}
\vspace{-0.3cm}
\end{table}

Table~\ref{tab:meta-profile-counts} reports mutually exclusive \emph{meta-feedback} categories, where each feedback instance is assigned to the exact combination of meta-clusters it contains. 
Helpfulness and task success is the most common meta-feedback (37.8\%), followed by correctness and truthfulness (24.2\%) and communication quality (11.5\%). Together, these three categories cover 73.5\% of all feedback, indicating that most dissatisfaction is driven by a single primary issue. Compound meta-feedback categories are less frequent but systematic. The leading combinations pair task success with communication quality (3.0\%), correctness with communication quality (2.9\%), and task success with both correctness and communication quality (2.5\%), suggesting that core failures often co-occur with presentation shortcomings. Cultural alignment appears both as a standalone category (1.8\%) and alongside communication quality (1.0\%). Safety is rare overall, but it most often appears with cultural alignment (1.3\% and 1.0\%), reinforcing that harmful content concerns in our setting are frequently tied to cultural appropriateness.

Table~\ref{tab:meta-cluster-marginals} reports the marginal prevalence of each meta-cluster across $N=988$ feedback instances, where a single instance may be assigned multiple clusters. Helpfulness and task success issues are the most common, appearing in 40.7\%, indicating that instruction adherence and direct task completion remain the primary sources of dissatisfaction. Correctness and truthfulness concerns are also frequent, affecting 29.9\%, which suggests that factual reliability is a major reason of negative feedback. Communication quality problems occur in 20.5\%, reflecting recurring issues with style, formatting, and fluency. Cultural alignment is less frequent but still notable at 68 cases (5.3\%), highlighting the importance of culturally appropriate responses in Arabic and Islamic contexts. Safety and harmlessness issues are comparatively rare (3.6\%), yet remain important given their potential severity.

\section{Experiments and Results}

\subsection{Experimental Setup}
We benchmark both open and closed models under multiple query input variations, including \textit{(i)} transcript using ASR vs. gold transcripts, and \textit{(ii)} raw audio. For ASR, as noted earlier, we use Fanar Aura and Gemini, since both have shown competitive performance for Arabic in prior work~\cite{alam2025spokennativqa}.
This setup allows us to evaluate both cascaded pipelines (ASR $\rightarrow$ LLM) and end-to-end audio-capable systems.

Among open models, we evaluate \textbf{ALLaM-7B}~\cite{bari2024allam} and \textbf{Fanar-2}~\cite{fanar2024} on transcriptions, and \textbf{Qwen2.5-Omni-3B}~\cite{xu2025qwen25omnitechnicalreport} directly on  audio. For closed models, we evaluate \textbf{GPT-5}~\cite{singh2025openai} on transcriptions, \textbf{GPT-4o Audio}~\cite{openai2023gpt4} directly on audio, and \textbf{Gemini-2.5 Pro}~\cite{team2023gemini} on both transcriptions and audio.

\subsection{Evaluations} 
%
\noindent
Since open-ended queries can have multiple valid responses, we assess response quality without gold references using a rubric-based LLM-as-a-judge method, which has been widely adopted in prior work \cite{zheng2023judging}. For each query, the judge receives the same input condition used by the evaluated model (i.e., audio for audio-based LLMs and transcription for text-based LLMs), without access to any reference. We use \textbf{Gemini 3 Pro} as the judge model.

We use rubric-based evaluation to assess model quality beyond correctness, capturing more nuanced dimensions of performance that have been adopted in recent work \cite{starace2025paperbench,guo2025beyond}. 
The rubric items include the following:
\begin{itemize}
    \item \textit{Intent precision} measures whether the response addresses the user’s intended task/request without drifting to irrelevant or mismatched content.
    \item \textit{Context awareness} measures whether the response correctly uses information referred to by the query and remains consistent with the dialogue state. 
    \item \textit{Specificity} measures whether the response provides concrete, actionable details rather than vague generalities.
    \item \textit{Depth and thoroughness} measures whether the response covers the key aspects needed to satisfy the request, including important conditions or steps when applicable.
    \item \textit{Grounding and honesty} measures whether the response avoids unsupported claims and communicates uncertainty or limitations when appropriate, rather than hallucinating.
    \item \textit{Format and language compliance} measures whether the response follows requested formatting constraints and uses the requested language and register consistently.
    \item \textit{Coherence} measures whether the response is logically organized, internally consistent, and easy to follow.
\end{itemize}

We report two aggregate metrics following prior rubric-based literature \cite{guo2025beyond}. Let $N$ be the total number of prompts in the test set, and let $R$ be the total number of applicable rubric criteria for a given prompt $i, i \in N$ . We define $\mathbbm{1}_{r_{ij}} \in \{0, 1\}$ as the binary indicator of whether the model's response to prompt $i$ satisfies rubric criterion $j$.

\noindent
\textbf{Average Pass Rate (APR).} 
A response counts as a pass only if it satisfies all required rubric checks. We define a binary pass indicator $p_i$ for query/prompt $i$ as the product of its individual rubric checks:
\begin{equation}
p_i = \prod_{j=1}^{R}\mathbbm{1}_{r_{ij}}
\end{equation}
APR is then the percentage of responses that pass completely across the whole test set:
\begin{equation}
\mathrm{APR} = \frac{1}{N}\sum_{i=1}^{N} p_i
\label{eq:apr}
\end{equation}

\noindent
\textbf{Average Rubric Score (ARS).}
ARS provides a more granular measure of partial success, computed as the mean pass rate across all individual rubric checks. For each query $i$, we compute the fraction of satisfied criteria:
\begin{equation}
s_i = \frac{1}{R}\sum_{j=1}^{R}\mathbbm{1}_{r_{ij}}
\label{eq:task_score}
\end{equation}
Equivalently, the overall ARS is the average of these per-query pass rates across the entire test set:
\begin{equation}
\mathrm{ARS} = \frac{1}{N}\sum_{i=1}^{N} s_i
\label{eq:ars}
\end{equation}

\subsection{Results}

\begin{table}[t]
\centering
\setlength{\tabcolsep}{1pt}
\scalebox{0.8}{
\begin{tabular}{l l r r}
\toprule
\textbf{Model} & \textbf{Input} & \textbf{APR (\%)} & \textbf{ARS (\%)} \\
\midrule

ALLaM-7B
    & Aura-ASR   & 23.71 & 64.58 \\
    & Gemini-ASR & \textbf{28.53} & \textbf{69.21} \\
    & Gold-trans & 28.14 & 69.50 \\
\cmidrule(lr){1-4}

Fanar-2
    & Aura-ASR   & 36.30 & 72.98 \\
    & Gemini-ASR & 40.61 & 76.49 \\
    & Gold-trans & \textbf{41.22} & \textbf{77.18} \\
\cmidrule(lr){1-4}

Gemini-2.5 Pro
    & Audio      & 82.01 & 91.97 \\
    & Aura-ASR   & 80.92 & 90.05 \\
    & Gemini-ASR & 89.01 & 94.09 \\
    & Gold-trans & \textbf{89.20} & \textbf{94.56} \\
\cmidrule(lr){1-4}
GPT-5
    & Aura-ASR   & 64.90 & 85.36 \\
    & Gemini-ASR & \textbf{73.25} & \textbf{89.57} \\
    & Gold-trans & 72.01 & 89.44 \\
\cmidrule(lr){1-4}

GPT-4o Audio
    & Audio      & 50.56 & 77.91 \\
\cmidrule(lr){1-4}

Qwen2.5-Omni-3B
    & Audio      & 3.30 & 40.34 \\
\bottomrule
\end{tabular}
}
\caption{APR and ARS results across models and input setups. APR: Average Pass Rate; ARS: Average Rubric Score.}
\label{tab:apr_ars}
\vspace{-0.3cm}
\end{table}


\begin{figure}[t]
\centering
\includegraphics[width=\linewidth]{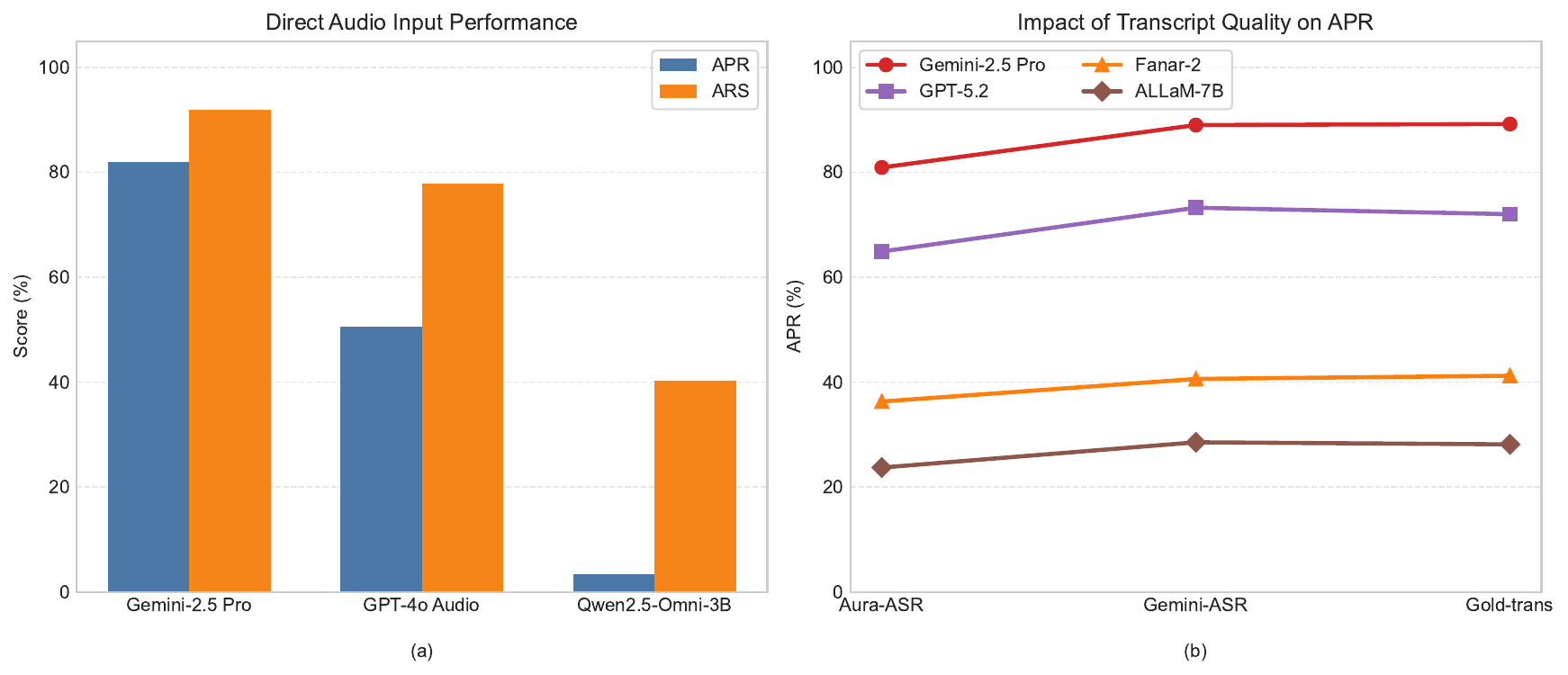}
\caption{Model performance metrics. \textit{(a)} Direct audio input performance across models. \textit{(b)} The impact of transcript quality on Average Pass Rate (APR) for cascaded models.}
\label{fig:performance_charts}
\vspace{-0.3cm}
\end{figure}

Table \ref{tab:apr_ars} and Figure \ref{fig:performance_charts} summarize the APR and ARS across all evaluated models and input conditions. Overall, closed-source models significantly outperform both open-source and audio-native counterparts, with Gemini-2.5 Pro achieving the highest overall scores. Based on the results, we observe three primary findings:

\begin{itemize}
    \item \textbf{Impact of ASR Errors:} Across all text-based models, gold transcripts consistently produce the best results. Performance typically decreases when ASR-generated transcripts are used, suggesting that transcription errors can propagate to downstream models and reduce overall system performance.
    \item \textbf{Direct Audio Performance Varies Drastically:} When processing direct audio without intermediate transcripts, Gemini-2.5 Pro achieves a highly impressive 82.01\% APR and 91.97\% ARS. Conversely, other models processing direct audio struggle significantly. GPT-4o Audio manages only moderate performance (50.56\% APR), while Qwen2.5-Omni-3B falls severely behind (3.30\% APR). This indicates that while end-to-end audio processing remains a major challenge for certain architectures, it can be highly effective in leading closed models.
    \item \textbf{The Open-Source Gap:} Among the open models, Fanar-2 consistently outperforms ALLaM-7B across all cascaded inputs. However, both underperform compared to the closed models by a wide margin. This gap is particularly evident in the APR metric, where the best open-source setup (Fanar-2 utilizing gold-transcritpions) scores 41.2\%, remaining well below the performance floor of the closed-source models.
\end{itemize}

\section{Discussion}


\noindent
\subsection{Effect of Input Modality and Transcript Quality}
\noindent
Table~\ref{tab:gemini_rubric_breakdown} details Gemini's performance across different input conditions and rubric dimensions. Performance generally improves as transcript quality increases, with Gemini-ASR closely approaching the results obtained using gold transcripts. 

When reasoning directly from audio, the most significant performance drops, compared to gold transcripts, occur in \textit{\textbf{depth}} (84.51\% vs.\ 91.67\%, a decrease of 7.16\%) and \textit{\textbf{specificity}} (85.42\% vs.\ 91.72\%, a 6.30\% drop), followed by \textit{\textbf{intent precision}} (a 3.06\% drop). Conversely, \textit{\textbf{context awareness}} (95.16\% vs.\ 94.84\%) and \textit{\textbf{coherence}} (99.41\% vs.\ 99.52\%) remain highly stable. This indicates that while the model maintains structural fluency and general contextual alignment when processing audio natively, it struggles to extract and synthesize the precise details required for deep, specific answers.

Providing transcripts does not uniformly improve performance, as results depend on transcript quality. Aura-ASR slightly underperforms direct audio (80.92\% vs.\ 82.01\% APR), while Gemini-ASR significantly improves performance to 89.01\% APR and 94.09\% ARS, closely matching gold transcripts (89.20\% APR and 94.56\% ARS). Differences between Gemini-ASR and gold are minimal across most metrics, indicating that high-quality ASR can nearly eliminate the gap to gold references, whereas lower-quality ASR can still degrade performance.


\begin{table}[t]
\centering
\setlength{\tabcolsep}{3pt}
\scalebox{0.80}{
\begin{tabular}{l r r r r}
\hline
\textbf{Rubric Metric} 
& \textbf{Audio} 
& \textbf{Aura-ASR} 
& \textbf{Gemini-ASR} 
& \textbf{Gold-trans} \\
\hline
Intent Precision      & 88.66 & 84.07 & 91.43 & 91.72 \\
Context Awareness     & 95.16 & 94.17 & 94.45 & 94.84 \\
Specificity           & 85.42 & 83.91 & 91.33 & 91.72 \\
Depth                 & 84.51 & 84.07 & 91.11 & 91.67 \\
Grounding             & 95.96 & 90.97 & 94.99 & 95.49 \\
Format Compliance     & 94.68 & 94.44 & 95.91 & 96.94 \\
Coherence             & 99.41 & 98.72 & 99.41 & 99.52 \\
\hline
APR                  & 82.01 & 80.92 & 89.01 & 89.20 \\
ARS                  & 91.97 & 90.05 & 94.09 & 94.56 \\
\hline
\end{tabular}
}
\caption{\textbf{Gemini performance across input variants under the judge rubric.} 
Input conditions include direct audio input, and transcripts such as Aura-ASR, Gemini-ASR, and gold. Numbers are in percentages (\%).
}
\label{tab:gemini_rubric_breakdown}
\vspace{-0.3cm}
\end{table}

\subsection{Effect of Dialect Variation}

\begin{table}[t]
\centering
\setlength{\tabcolsep}{1pt}
\scalebox{0.8}{
\begin{tabular}{l l r r}
\toprule
\textbf{Variant} & \textbf{Country} & \textbf{APR (\%)} & \textbf{ARS (\%)} \\
\midrule

Audio
& DZ & 66.28 & 85.05 \\
& EG & 79.91 & 91.24 \\
& SD & 84.88 & 92.94 \\
& SY & 83.72 & 91.69 \\
\cmidrule(lr){1-4}

Fanar-ASR
& DZ & 66.28 & 84.22 \\
& EG & 79.26 & 89.01 \\
& SD & 79.19 & 89.93 \\
& SY & 76.74 & 88.70 \\
\cmidrule(lr){1-4}

Gemini-ASR
& DZ & 80.00 & 88.07 \\
& EG & 87.48 & 93.13 \\
& SD & 91.12 & 95.27 \\
& SY & 86.05 & 92.69 \\
\cmidrule(lr){1-4}

Gold-trans
& DZ & 76.47 & 86.55 \\
& EG & 85.37 & 92.18 \\
& SD & 94.80 & 97.36 \\
& SY & 93.02 & 97.01 \\
\bottomrule

\end{tabular}
}
\caption{Gemini performance across dialects and input variants.}
\label{tab:gemini_dialect_impact}
\vspace{-0.3cm}
\end{table}

Table~\ref{tab:gemini_dialect_impact} shows clear performance variation across dialects, with Algerian Arabic (DZ) consistently being the most challenging and Egyptian (EG), Sudanese (SD), and Syrian (SY) performing better overall. Under direct audio input, performance is lowest for DZ (66.28\% APR), while SD and SY achieve the highest scores.

Introducing ASR transcripts does not uniformly improve performance. Fanar-ASR often leads to similar or slightly lower results compared to audio, whereas Gemini-ASR provides consistent gains across all dialects, improving APR from 66.28\% to 80.00\% for DZ and from 79.91\% to 87.48\% for EG. Postedited transcripts yield the strongest overall results, particularly for SD (94.80\% APR) and SY (93.02\% APR). Overall, the results suggest that dialectal variation strongly affects performance, and that high-quality transcription is key to reducing this gap, especially for lower-resource dialects such as DZ.

\subsection{Transcription Uncertainty and User Dissatisfaction}
Figure~\ref{fig:asr_sim_likes_dislikes} shows transcription difficulty to real user reactions. Prompts that receive dislikes exhibit lower agreement between Fanar and Gemini transcriptions than prompts that receive likes, indicating that the disliked subset contains a higher proportion of acoustically or linguistically challenging utterances. This supports two practical uses of multi-ASR signals in spoken LLM systems (both cascaded and end-to-end audio-native pipeline). First, inter-ASR agreement can serve as a lightweight indicator for selecting samples that are likely to require post-editing when constructing reference transcripts. Second, at runtime it can be used to detect high-uncertainty turns where clarification or confirmation is preferable to a confident but potentially off-target response.

\subsection{What Users Dislike in Spoken LLM interactions}
Tables~\ref{tab:meta-profile-counts} and \ref{tab:meta-cluster-marginals} show that most dissatisfaction is driven by a small set of failure modes. The dominant marginal dimensions are \emph{helpfulness/task success} and \emph{correctness/truthfulness}, while \emph{communication quality} is a frequent secondary issue. This is consistent with the rubric trends in Table~\ref{tab:gemini_rubric_breakdown}, where audio input reduces \textit{specificity} and \textit{depth}, properties that are closely related to perceived instruction following and completeness. Cultural alignment and safety feedback occur less frequently, but remain important because of their potential severity and their tendency to co-occur with other issues.

\section{Conclusion}
In this paper, we introduced \textsc{WASIL}, to our knowledge the first in-the-wild dataset of Arabic spoken interactions with LLMs, designed to capture realistic conversational conditions under dialect variation and speech-driven input noise. The dataset includes post-edited transcriptions, user feedback (like and dislike, with fine-grained categories for disliked responses), and dialect and answerability annotations. To reduce the cost of producing gold transcriptions at scale, we propose a multi-ASR similarity-based approach that prioritizes samples for human post-editing. Our experiments show that performance under audio and ASR inputs can diverge substantially from gold-transcription settings, and that downstream robustness is not fully explained by WER alone. Instead, dialect coverage and a model's ability to handle dialectal lexical and syntactic variation play an important role in how transcription errors translate into task failures.
\dataset{} enables several research directions. First, it supports principled studies of error propagation from ASR to LLM reasoning and generation, including controlled comparisons across multiple transcripts for the same utterance. Second, it provides a foundation for preference-based alignment using like and dislike signals with audio-grounded judgments, which better reflect user intent in spoken interactions. Finally, the dataset can facilitate future work on culturally aware Arabic assistants by enabling targeted evaluation of culturally sensitive failures. 



\noindent
\textbf{Limitations.}
\textsc{WASIL} captures realistic spoken interactions, though it reflects the user population available in our collection pipeline, which may under-represent certain dialects and use cases. The like and dislike signals provide useful indications of user satisfaction, although they do not fully capture all aspects of response quality and may emphasize more salient failure cases or include limited rationales. Our rubric-based evaluation does not rely on gold references for open-ended queries, which allows broader coverage of responses. At the same time, results can be influenced by the choice of judge model and rubric interpretation, and findings may continue to evolve as ASR and LLM systems improve.

\noindent
\textbf{Broader Impact.}
\textsc{WASIL} can support the development of dialect-inclusive Arabic speech assistants and allows researchers to evaluate exactly how ASR errors degrade LLM responses. While this fine-grained feedback helps improve system reliability, the underlying speech data presents inherent privacy, re-identification, and bias risks. Consequently, we advocate for responsible governance and intended use of this dataset.

\section{Generative AI Use Disclosure}
We used generative AI tools to support writing, specifically, we employed an LLM to help refine wording, and improve clarity. All methodological choices, experimental design, reported results, and interpretations were defined by the authors.

\bibliographystyle{IEEEtran}
\bibliography{bibliography}

\clearpage
\newpage

\end{document}